\documentstyle[11pt,aaspp4,epsf]{article}

\newcommand{\be}{\begin{equation}}
\newcommand{\ee}{\end{equation}}
\def\nh{n_{H,21}}

\def\ros{{\sl ROSAT}}
\def\chan{{\sl Chandra}}
\def\asca{{\sl ASCA}}

\def\pdot{\dot{P}}
\def\fdot{\dot{f}}

\def\edot{\dot{E}}

\def\nh{n_{\rm H,21}}

\def\ns{1E~1207.4--5209}
\def\Z{Z^2_1}
\def\Zm{Z^2_{1,{\rm max}}}

\begin{document}

\title{Discovery of 424~ms pulsations from the
radio-quiet neutron star in the PKS 1209--52 supernova remnant}
\author{V.~E.~Zavlin\footnote{
Max-Planck-Institut f\"ur Extraterrestrische Physik, D-85740
Garching, Germany; zavlin@xray.mpe.mpg.de}, 
G.~G.~Pavlov\footnote{
The Pennsylvania State University, 525 Davey Lab,
University Park, PA 16802, USA; pavlov@astro.psu.edu}, 
D.~Sanwal$^2$, and J.~Tr\"umper$^1$}
%%%%%%%%%%%%%%%%%%%%%%%%%%%%%%%%%%%%%%%%%%%%%%%%%%%%%%%%%%%%%%%%%%%%%%
\begin{abstract}
The central source of the supernova remnant PKS 1209--52
was observed with the Advanced CCD Imaging Spectrometer 
aboard {\sl Chandra} X-ray observatory on 2000 January 6--7.
The use of the Continuos Clocking mode allowed us to perform
the timing analysis of the data with time resolution of 
2.85~ms and to find a period $P=0.42412927\pm 2.3\times 10^{-7}$~s.
The detection of this short period proves that the source
is a neutron star. It may be either an active pulsar with
unfavorably directed radio beam or a truly radio-silent
neutron star whose X-ray pulsations are caused by a nonuniform distribution
of surface temperature.
To infer the actual properties of this neutron star,
the period derivative should be measured.
\end{abstract}
%%%%%%%%%%%%%%%%%%%%%%%%%%%%%%%%%%%%%%%%%%%%%%%%%%%%%%%%%%%%%%%%%%%%%
%%%%%%%%%%%%%%%%%%%%%%%%%%%%%%%%%%%%%%%%%%%%%%%%%%%%%%%%%%%%%%%%%%%%%%
\section{Introduction}
Radio-quiet compact central objects (CCOs)
of supernova remnants (SNRs) have emerged recently
as a separate class of X-ray sources 
(see, e.g., Kaspi 2000, for a review). 
The nature of these
sources, which are expected to be either neutron stars (NSs)
or black holes formed in supernova explosions,
remains enigmatic. They are characterized by soft,
apparently thermal,
X-ray spectra and a lack of visible pulsar activity
and optical counterparts. 
One of the most important properties, which potentially
allows one to elucidate the nature of these sources, is
periodicity of their radiation. Some of these sources
show periods in the range of 6--12 s, much longer than
those of typical radio pulsars. These object form a subclass
of anomalous X-ray pulsars (e.g., Gotthelf \& Vasisht 2000),
which have been interpreted as magnetars (NSs with superstrong
magnetic fields --- see Thompson 2000).
A putative period of 75 ms was proposed by 
Pavlov, Zavlin, \& Tr\"umper 
(1999) for the Puppis A CCO; however, 
a low significance of that detection requires it to be confirmed
by other observations. An unusually long
period of 6 hours has been reported for the  RCW 103 CCO
(Garmire et al.~2000a), the origin of this periodicity is unclear.
No periodicity has been found for the recently discovered
CCO of the Cas A SNR (Pavlov et al.~2000; Chakrabarty et al.~2000).

In this paper we report detection of a period of 
\ns, the CCO of PKS 1209--52, a barrel-shaped radio, X-ray, and
optical SNR (also known as G296.5+10.0). From
the analysis of radio and optical observations of this SNR,
Roger et al.~(1988) estimated its age $\sim 7000$~yr,
with an uncertainty of a factor of 3.
A recent estimate of the distance to PKS 1209--52, 
$d=2.1^{+1.8}_{-0.8}$~kpc, is given by Giacani et al.~(2000). 
Estimates of the interstellar hydrogen column density from the
radio, optical, and UV data yield $\nh\equiv n_H/(10^{21}~{\rm cm}^{-2})
\sim 1.0$--1.8 (see Kellet et al.~1987;  Roger et al.~1988;
Giacani et al.~2000),
consistent with a distance $d_2 \equiv d/(2~{\rm kpc})\sim 1$. 

The point source \ns\ was discovered
with the {\sl Einstein} observatory (Helfand \& Becker 1984),
$6'$ off-center the $81'$ diameter SNR. 
Mereghetti, Bignami \& Caraveo
(1996) and Vasisht et al.~(1997)
showed that the \ros\ and \asca\ spectra of \ns\
can be interpreted as blackbody emission of $T\simeq 3$~MK
from an area with radius $R\simeq 1.5~d_2$~km
and suggested that this radiation comes from hot spots on NS surface.
The spots may be heated either by
dissipative heating in the NS interior or by the bombardment
of polar caps by relativistic particles from the NS magnetosphere
if \ns\ is an active pulsar.
However, the former hypothesis can hardly explain the small
sizes of the hot spots, even with  allowance for large anisotropy
of thermal conductivity of the magnetized NS crust.
The latter heating mechanism is also in doubt because of
the absence of radio and $\gamma$-ray emission from \ns.

 From observations at 4.8 GHz, Mereghetti et al.~(1996) found an upper limit
of $\sim 0.1$~mJy on the radio flux from \ns.
They also set a deep limit of $V > 25$ for an optical counterpart
in the {\sl Einstein} HRI error circle, which supports the hypothesis that
\ns\ is indeed an isolated NS.

Zavlin, Pavlov \& Tr\"umper (1998) reanalyzed the \ros\ and
\asca\ data, fitting the observed spectra with NS atmosphere
models. They have shown that the hydrogen atmosphere fits yield more
realistic parameters of the NS and the intervening hydrogen column than
the traditional blackbody fit. In particular, for a NS of mass $1.4~M_\odot$
and radius 10~km, they obtained a NS surface temperature
$T_{\rm eff}=(1.4$--$1.9)$~MK and distance $d=1.6$--3.3 kpc,
versus $T=(4.2$--$4.6)$~MK
and implausibly large $d=11$--13~kpc
for the blackbody fit, at a 90\% confidence level.
The hydrogen column density inferred from the atmosphere fits,
$\nh = 0.7$--2.2,
agrees fairly well with independent estimates obtained from UV
observations of nearby stars, radio data, and X-ray spectrum
of the shell of the supernova remnant, whereas the blackbody and
power-law fits give considerably lower and greater values,
$\nh=0.2$--0.4 and $\nh=5.2$--7.0,
respectively. The NS surface temperature inferred from the atmosphere fits
is consistent with standard NS cooling models.

All the previous observations failed to detect pulsations of X-ray
radiation from \ns. An upper limit of 18\% (at a 95\% confidence level)
on flux modulation was set by Mereghetti et al.~(1996)
 from the analysis of the \ros\ PSPC data. 
Due to its higher sensitivity and the possibility of continuous
observations, the {\sl Chandra} X-ray Observatory 
is much more capable to search for periodicity of X-ray sources.
We have employed this capability to search for the period of \ns.
The observation and data reduction are described in \S 2,
the timing analysis of the data is presented in \S 3, and
some implementations are briefly discussed in \S 4.

%%%%%%%%%%%%%%%%%%%%%%%%%%%%%%%%%%%%%%%%%%%%%%%%%%%%%%%%%%%%%%%%%
\section{Observation and data reduction}
\ns\ was observed with \chan\ on 2000 January 6-7 
with the spectroscopic array of the Advanced CCD Imaging Spectrometer
(ACIS --- see Garmire et al.~2000b)
in the Continuous Clocking (CC) mode. 
This mode provides the highest time resolution of
2.85~ms available with ACIS by means of sacrificing
spatial resolution in one dimension.
The source was imaged on the back-illuminated chip S3.
The total duration
of the observation was 32.6~ks.
Time history of detected events reveals that there were three
relatively short time intervals with very strongly, by an order of
magnitude,  increased background
(background ``flares'') distributed over the whole detector.
To mitigate the contamination of the source by the flares,
we excluded these intervals from further analysis,
which resulted in the effective exposure of 29,283~s.

The event arrival times were not properly corrected
for the satellite wobbling (dither) by
the standard pipe-line processing.
 From our analysis (Sanwal et al.~2000) of \chan\  data on PSR 1055--52,
observed in the same mode, we have been aware that the lack
of this correction results in false periodicities (side peaks
in the power spectrum) of a pulsar.
Therefore, we corrected the event arrival times $t$ 
in the event file making use of the formula
(Glenn Allen, private communication):
\be
t_{\rm corr}=t+c_1~\left[(\alpha-\alpha_{\rm m})~\sin\zeta~\cos\delta_{\rm m} -
(\delta-\delta_{\rm m})~\cos\zeta\right]~,
\ee
where $c_1=c_2\, \delta t$,
$c_2=3600''/0\farcs 4919 = 7318.5605$ is the
scaling coefficient transforming degrees to the detector 
pixels, $\delta t=2.85$~ms is the integration time in the CC mode,
$\zeta$ is the roll angle, and
$\alpha-\alpha_m$ and $\delta-\delta_m$ (in degrees) are the
deviations of right ascension and declination
from their median values (calculated with the use of the
aspect solution file).  We have checked that this correction 
removes the artificial  periodicities in the data on PSR 1055--52.

The 1D image of \ns\ in ``sky pixels'' (Fig.~1) is
consistent with the ACIS PSF.
To obtain the source count rate, we extracted
23,337 source+background counts from a 1D segment
of 8 pixel length centered at the source position
(10 pixels contain 23,602 counts).
The background  was taken from similar segments
adjacent to the 1D source aperture.
Subtracting the background, we find the source countrate of
$0.76\pm0.01$~s$^{-1}$. 
A preliminary analysis of the source spectrum allows us
to estimate the observed source energy flux $f_x=2.2\times 10^{-12}$
erg~cm$^{-2}$~s$^{-1}$, in the 0.1--5.0~keV range.
The corresponding (unabsorbed) luminosity is $L_x\approx 1.3\times 10^{33}
d_2^2 $ erg~s$^{-1}$, and the bolometric luminosity 
(corrected for the gravitational redshift at the surface
of NS with $M=1.4M_\odot$, $R=10$ km) is 
$L_x\approx 5\times 10^{33} d_2^2 $ erg~s$^{-1}$,
in agreement with the estimates obtained by Zavlin et al.~(1998).

%%%%%%%%%%%%%%%%%%%%%%%%%%%%%%%%%%%%%%%%%%%%%%%%%%%%%%%%%%%%%%%%%%%%%%
\section{Timing analysis}
Previous $ROSAT$ and $ASCA$ observations of \ns\ have not revealed
pulsations of X-ray radiation from this object. This may be explained
by small amount of counts collected in those observations. Moreover,
those data were spread over very long time intervals, that always 
heavily complicates searching for weak pulsations.

One of the main advantages of the \chan\ observation is that a large number
of the source counts (about 10 times the number of counts
previously detected) has been acquired in a short time span.
The spectral analysis shows that the background 
exceeds the source radiation at energies above 3~keV.
Therefore, for
the timing analysis we chose 22,535 counts with energies below 3~keV
from a 6-pixel segment centered at the \ns\ position. 
Of these counts, 
about 98\% were estimated to belong to the source.
To search for pulsations, we used 
the well-known $\Z$ (Rayleigh) test
(Buccheri et al.~1983), which is expected to be optimal to search for
smooth pulsations.  We ran the test in
the 0.01--100~Hz frequency range with a step 
${\delta}f=3$ $\mu$Hz.
The oversampling by a factor of 10 (compared to the expected widths
of $\sim 1/T$ of the $\Z$ peaks, where $T=32.6$~ks is the observational span) 
was chosen 
to resolve separate $\Z$ peaks; it guarantees that the peak corresponding
to a periodic signal would not be missed.
The number of statistically independent trials in the chosen
frequency range can be estimated as
${\cal N}=100~{\rm Hz}\times T=3.26\times 10^6$.
The test yields only one high peak of $\Zm=65.0$ at
the frequency $f_*=2.357769$~Hz (see Fig.~2).
Since the variable $\Z$ has a probability density function
equal to that of a $\chi^2$ with two degrees of freedom,
the probability to obtain a noise peak of a given height in one trial is
${\exp}(-\Z/2)$. This means that the probability to obtain {\it by chance}
a peak of $\Z=65.0$ in ${\cal N}$ independent trials is 
$\rho={\cal N}{\exp}(-\Zm/2)=2.5\times 10^{-8}$, 
that corresponds to a detection of pulsations at a
confidence level of $C=(1-\rho)\times 100\%=99.9999975\%$,
or $5.5\sigma$. The next heighest peak is  $\Z=32.0$. The probability
to obtain such a peak by chance in ${\cal N}$ trials is 35\%.

To be sure that the pulsations found at the frequency $f_*$ are
not associated with instrumental effects, we ran the $\Z$ test at $f=f_*$
on the same amounts of counts extracted from a few background regions
and obtained maximum $\Z$ equal to 5. We also varied the size of the segment
for source counts extraction and found that the initial choice of 6-pixel 
length was optimal to produce the maximum $\Z$ value.
We found that contributions
from higher harmonics to the value of $Z^2$ are insignificant:
$Z^2_2=65.1$ and $Z^2_3=65.2$. 

To evaluate the pulsation frequency more precisely
and find its uncertainty, we employed the 
method suggested by Gregory \& Loredo (1996; GL hereafter), 
based on the Bayesian formalism.
The method uses the phase-averaged epoch-folding
algorithm to calculate a frequency-dependent odds ratio, $O_m(f)$,
which specifies
how the data favor a periodic model of a given frequency $f$ with
$m$ phase bins over the unpulsed model (see Gregory \& Loredo 1992 for
details of odds ratio computations). 
To weaken the dependence on number of bins,
GL suggested to use the odds ratio 
$O_{\rm per}(f)=\sum_{m=2}^{m_{\rm max}} O_m(f)$, with a characteristic
number of $m_{\rm max}=10$--15. The distribution
of probability for a signal to be periodic, with frequency $f$
in a chosen frequency range $(f_1,f_2)$, can be written as
\be
p(f)=\frac{O_{\rm per}(f)}{f}
\left[(1+O_{\rm per}^*)\, \ln\frac{f_2}{f_1}\right]^{-1}~,
\ee
where 
\be
O_{\rm per}^*=
\left(\ln\frac{f_2}{f_1}\right)^{-1} \int_{f_1}^{f_2}\frac{{\rm d}f}
{f}O_{\rm per}(f)~,
\ee
and the probability for the signal to be periodic in $(f_1,f_2)$ is
$O_{\rm per}^*/(1+O_{\rm per}^*)$.
Using PSR 0540--69 as an example, GL
have demonstrated that this method allows
one to determine the pulsation frequency with much
higher accuracy than it can be done 
with more traditional methods. The most probable frequency $f_0$ is
given by the maximum of $O_{\rm per}(f)$ in $(f_1,f_2)$.
The uncertainty $\delta f$ at a given confidence 
level $C=\tilde{\rho}\times
100\%$ can be calculated from the following equation 
\be
\int_{f_0-{\delta}f/2}^{f_0+{\delta}f/2} p(f)\, {\rm d}f
=\tilde{\rho}~.
\ee

We implemented the GL method, taking $m_{\rm max}=12$
and $f_2-f_* = f_*-f_1 = 200$ $\mu$Hz.
The frequency dependence of odds ratio is shown in Figure 2. 
The maximum value, $O_{\rm per}^{\rm max}=6.2\times 10^8$,
is  at $f=f_0=2.3577717$~Hz ($f_0-f_*\simeq 3~\mu{\rm Hz}
\simeq 0.1/T$).
The uncertainty of $f_0$ at 68\%, 90\%, and 95\% confidence levels
is ${\delta}f=1.3$, 2.5, and 3.5 $\mu$Hz, respectively.
This estimate is practically independent of choice of frequency
range if $\Gamma \ll f_2-f_1 \ll f_*$, where 
$\Gamma\simeq 3$ $\mu$Hz
is a characteristic width of
the peak of the odds ratio (see Fig.~2).
Thus, we finally derive the frequency
and the period of the detected pulsations 
\be
f_0=2.3577717\pm 1.3\times 10^{-6}~{\rm Hz},
\ee
\be
P_0=0.42412927\pm 2.3\times 10^{-7}~{\rm s}~,
\ee
at the epoch of 51549.630051303 MJD (TDB).

The light curve extracted at $f=f_0$ (Fig.~3)
reveals one broad pulse per period with intrinsic source pulsed fraction of
$f_p=9\pm 2\%$. Assuming
that the detected signal is sinusoidal, we obtain an estimate on
the pulsed fraction $f_p=(2\Zm/N)^{1/2}=7.6\%$
(where $N$ is the number of counts), in a fair agreement with the value
calculated from the extracted light curve.
 Figure 3 indicates that the shape of the light curve
may vary slightly with photon energy.

%%%%%%%%%%%%%%%%%%%%%%%%%%%%%%%%%%%%%%%%%%%%%%%%%%%%%%%
\section{Discussion}
%%%%%%%%%%%%%%%%%%%%%%%%%%%%%%%%%%%%%%%%%%%%%%%%%%%%%%%
The detection of the period, $P\simeq 424$~ms,
proves that \ns, the central compact object
of the PKS 1209--52 SNR, is the neutron star.

Let us assume that \ns\ is
an active pulsar of an age $\tau$
within a range of 2--20 kyr (estimated limits of the SNR age).
Then, for a braking index of 2.5,
we should expect $\fdot\sim f/(1.5 \tau) \sim 
(2.5$--$25)\times 10^{-12}$~Hz~s$^{-1}$, $\pdot=
(4.5$--$45)\times 10^{-13}$~s~s$^{-1}$. A crude
estimate for the pulsar magnetic field
(at the magnetic pole) would be $B\sim
6.4\times 10^{19}(P\pdot)^{1/2} \sim 
(3$--$9)\times 10^{13}$~G,
close to the critical value $B_{\rm cr}=4.4\times 10^{13}$ G,
above which nonlinear QED effects can affect the processes 
in the pulsar magnetosphere.
The pulsar spin-down luminosity (rotation energy loss) would be $\edot = 
(2.3$--$23)\times 10^{35} I_{45}$~erg~s$^{-1}$. 
This $\edot$ may be high enough to power a compact
synchrotron nebula --- for instance, Gaensler et al.~(1998)
found a radio nebula with a radius of $0.3(d/7~{\rm kpc})$~pc
around a 100-kyr old 
PSR B0906--49, whose spin-down luminosity is $4.9\times
10^{35}$~erg~s$^{-1}$. A nebula of similar size would have
an angular radius of $\sim 30''$ at a distance of 2~kpc.
The 1D image of \ns\ puts an upper limit of $3''$--$4''$
on the radius of a compact nebula 
around the pulsar.

If \ns\ is an ordinary
radio pulsar whose radio-quiet nature 
is due to an unfavorable orientation of the pulsar beam,
the pulsar could be detected in deep radio
observations at low frequencies, where radio beams are broader.
On the other hand,
the lack of manifestations of pulsar
activity may indicate
that \ns\ is {\it not} an active
pulsar, e.g., because a very high magnetic field
can inhibit the cascade processes in the pulsar's 
acceleration zone (Baring \& Harding 1997). In this case,
the most natural explanation of the observed pulsations
would be anisotropy of temperature distribution caused
by anisotropic heat conduction in a superstrong magnetic
field (Greenstein \& Hartke 1983). 
This hypothesis can be
verified by a phase-dependent spectral analysis which
will be done on the same data when the ACIS response
is known with better precision.
Particularly useful for elucidating
the nature of the NS would be measuring of
the period derivative, 
which requires at least one more observation.
For instance, a similar ACIS observation in CC mode,
taken a year later, would
allow us to determine the frequency with an accuracy of
$\sim 2\times 10^{-6}$ Hz
 and detect a frequency derivative if it exceeds
$\sim 2\times 10^{-13}$
Hz~s$^{-1}$, which is substantially smaller than the above
estimates.

\acknowledgements
Our thanks are due to Glenn Allen, who provided the algorithm to
correct the event times for dither.
We are grateful to Gordon Garmire,
John Nousek, Leisa Townsley and George Chartas
for the useful advice on the analysis of ACIS data.
This work was partly supported by SAO grant GO0-1012X.

{}
%%%%%%%%%%%%%%%%%%%%%%%%%%%%%%%%
\newpage
\figcaption{Distribution of detected counts over pixels in
1D image in the vicinity of \ns. One sky pixel is equal to $0\farcs 492$}

\figcaption{Power spectrum (upper panel)
and frequency dependence of odds ratio
(lower panel) around 
$f_*=2.357769$.}

\figcaption{Light curves extracted in different energy ranges
period $P_0$.}

%%%%%%%%%%%%%%%%%%%%%%%

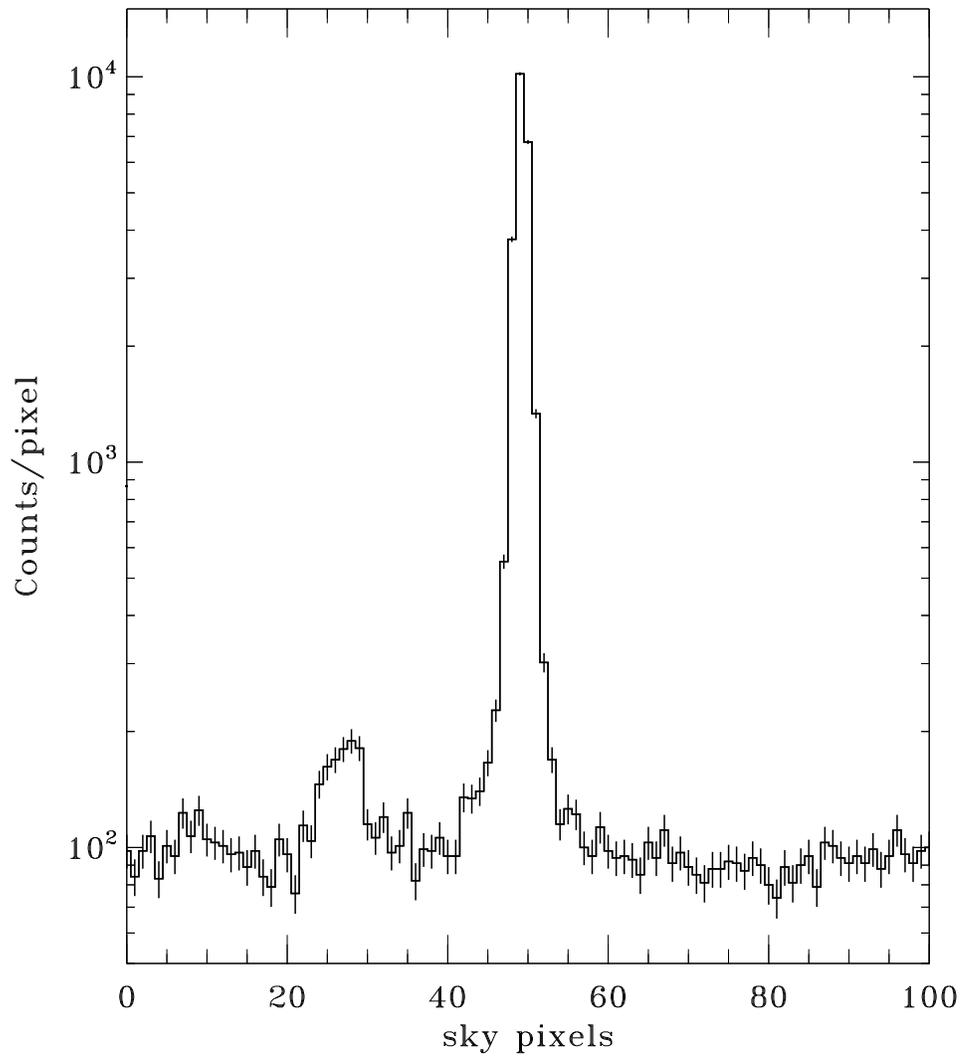
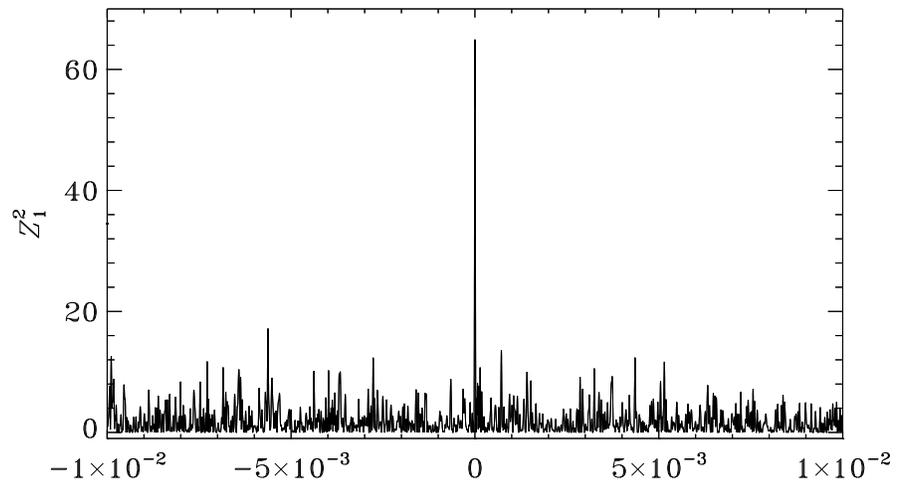
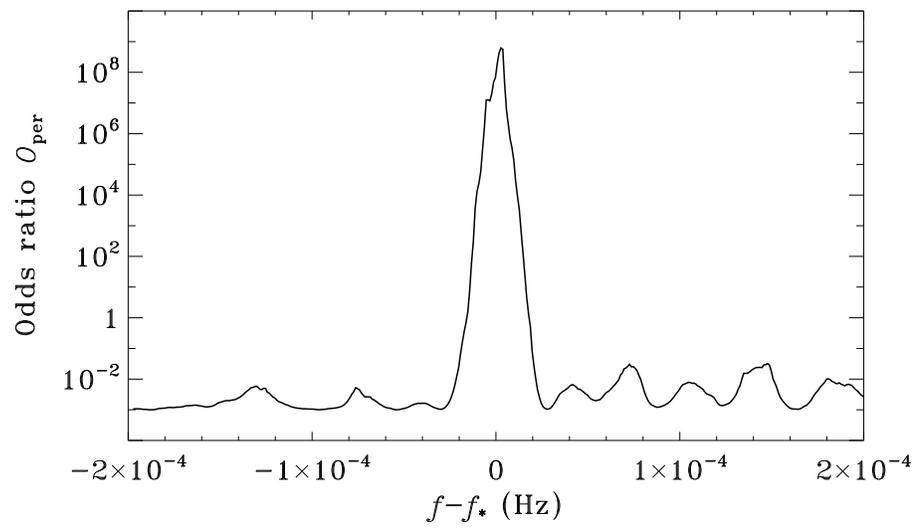
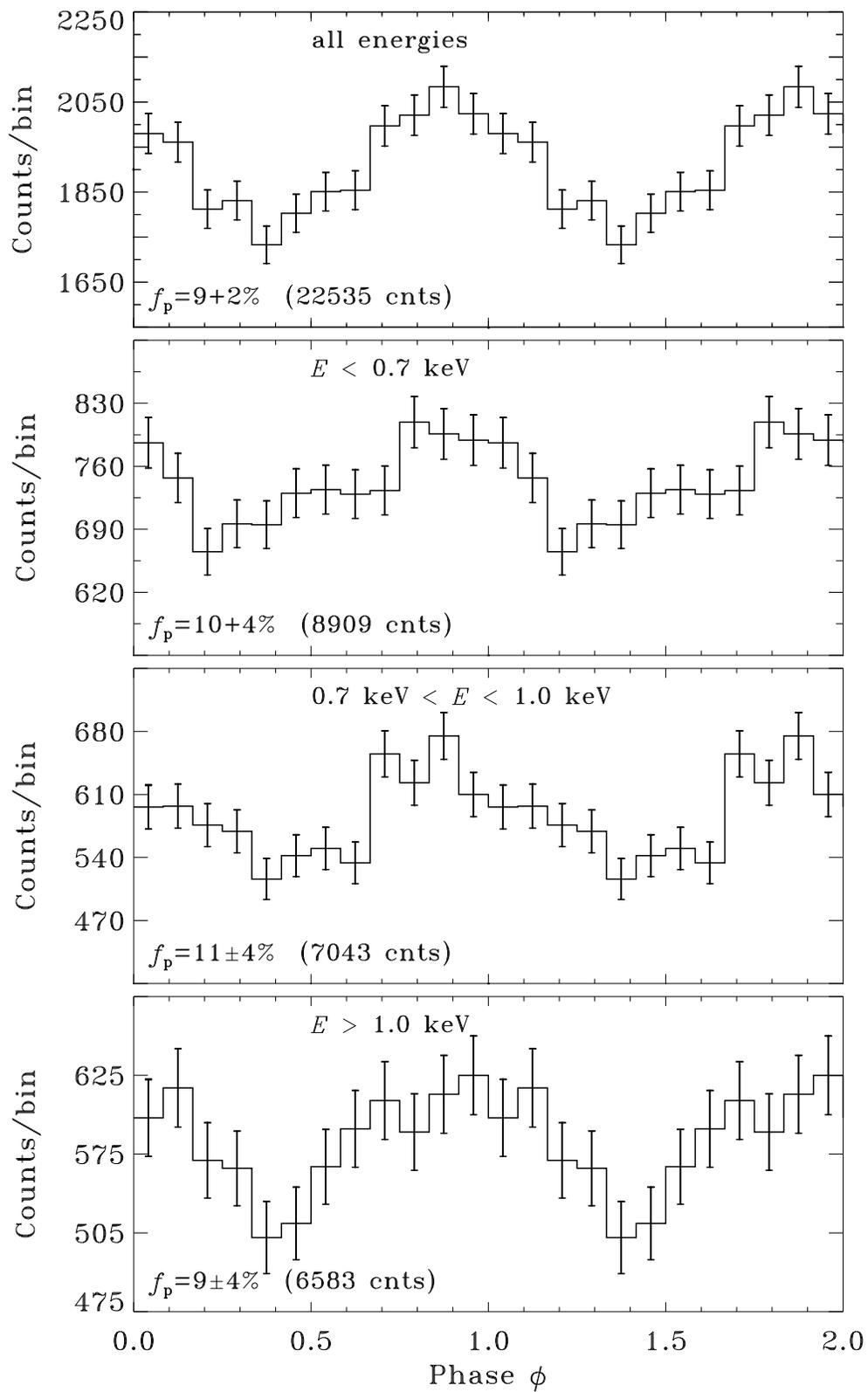
\end{document}